\documentclass[reqno,11pt]{amsart}
\usepackage{graphicx}
\usepackage{slashed}
\usepackage{amscd}
\usepackage{tensor}
\usepackage{amssymb}
\usepackage{esint}
\usepackage{enumitem}
\usepackage{accents}
\usepackage[usenames,dvipsnames]{pstricks}
\usepackage{pst-grad} % For gradients
\usepackage{pst-plot} % For axes
\usepackage[mathscr]{eucal}
\numberwithin{equation}{section}
\allowdisplaybreaks[1]

\title[Causal fermion systems]{Causal Fermion Systems: \\ A Primer for
Lorentzian Geometers}
\author[F.\ Finster]{Felix Finster \\ \\ September 2017}
\address{Fakult\"at f\"ur Mathematik \\ Universit\"at Regensburg \\ D-93040 Regensburg \\ Germany}
\email{finster@ur.de}

\newtheorem{Def}{Definition}[section]
\newtheorem{Thm}[Def]{Theorem}

\newcommand{\Thanks}{\vspace*{.5em} \noindent \thanks}
\newcommand{\beq}{\begin{equation}}
\newcommand{\eeq}{\end{equation}}
\newcommand{\Proof}{\begin{proof}}
\newcommand{\QED}{\end{proof} \noindent}

\newcommand{\la}{\langle}
\newcommand{\ra}{\rangle}

\newcommand{\Sl}{\mbox{$\prec \!\!$ \nolinebreak}}
\newcommand{\Sr}{\mbox{\nolinebreak $\succ$}}

\newcommand{\C}{\mathbb{C}}
\newcommand{\R}{\mathbb{R}}
\newcommand{\1}{\mbox{\rm 1 \hspace{-1.05 em} 1}}

\newcommand{\N}{\mathbb{N}}

\DeclareMathOperator{\tr}{tr}
\renewcommand{\O}{{\mathscr{O}}}
\renewcommand{\L}{{\mathcal{L}}}

\newcommand{\U}{\text{\rm{U}}}

\newcommand{\Cisc}{C^\infty_{\text{\rm{sc}}}}

\newcommand{\Dir}{{\mathcal{D}}}
\DeclareMathOperator{\supp}{supp}
\renewcommand{\H}{\mathscr{H}}

\newcommand{\Lin}{\text{\rm{L}}}
\newcommand{\F}{{\mathscr{F}}}

\DeclareMathOperator{\Symm}{\mbox{\rm{Symm}}}
\newcommand{\nablaLC}{\nabla^\text{\tiny{\tt{LC}}}}
\newcommand{\DLC}{D^\text{\tiny{\tt{LC}}}}

\newcommand{\scrM}{\mycal M}
\newcommand{\scrN}{\mycal N}

\newcommand{\itemD}{\item[{\raisebox{0.125em}{\tiny $\blacktriangleright$}}]}

\DeclareFontFamily{OT1}{rsfso}{}
\DeclareFontShape{OT1}{rsfso}{m}{n}{ <-7> rsfso5 <7-10> rsfso7 <10-> rsfso10}{}
\DeclareMathAlphabet{\mycal}{OT1}{rsfso}{m}{n}

\newcommand{\cliff}{\!\cdot\!}

\newcommand{\Cl}{{\mathscr{C}}\ell}
\newcommand{\nn}{\mathcal{N}}
\DeclareMathOperator{\Texp}{Texp}
\DeclareMathOperator{\s}{\text{\rm{scal}}}

\newcommand{\con}{\text{\rm{con}}}

\begin{document}
\maketitle

\begin{abstract}
We give a brief introduction to causal fermion systems
with a focus on the geometric structures in space-time.
\end{abstract}

\tableofcontents

In general relativity, space-time is described by a Lorentzian manifold.
Although this description has been highly successful, it is generally believed
that in order to reconcile general relativity with quantum theory,
the mathematical structure of space-time should be modified on a
microscopic scale (typically thought of as the Planck scale $\approx 10^{-35}\:
\text{meters}$). Apart from having far-reaching physical implications,
such a modification is also of purely mathematical interest because
going beyond the usual space-time continuum makes it possible to
model non-smooth structures in space-time and of space-time itself
which cannot be described (or at least are difficult to describe)
on a Lorentzian manifold.

Causal fermion systems are a recent physical theory based on a novel
mathematical model of space-time.
The present paper is an introduction to causal fermion systems
with a focus on the resulting geometric structures.
It is addressed to readers who are familiar with Lorentzian geometry
and the Dirac equation. 
More basic introductions in Minkowski space are given in the textbooks~\cite[Chapter~1]{cfs}
or~\cite{intro}. Also, we do not explain the broader physical picture
but refer instead to the non-technical survey~\cite{dice2014}.
Moreover, in~\cite{topology} some of the concepts are introduced
starting from Riemannian geometry.
For the physical applications, we refer to~\cite[Chapters~3-5]{cfs} and~\cite{qft}.
The relation to other approaches is explained in~\cite[\S3.4.6]{cfs}.

The paper is organized as follows. We first explain how one gets from Lorentzian geometry to
the setting of causal fermion systems (Section~\ref{sectoCFS}).
Then a few structures of an abstract causal fermion system are reviewed
(Section~\ref{secabstract}).
In Section~\ref{seclqg} we turn attention to the geometric structures
and explain how Lorentzian geometry is recovered as a limiting case.
In Section~\ref{secbeyond} it is explained why and specified how
the geometric structures of a causal fermion system go beyond Lorentzian geometry.
We focus on two aspects: First, space-time need not be smooth and the
structures therein need not be regular (Section~\ref{secnoreg}).
Second, causal fermion systems allow for the description
of collections of many different space-times which mutually
interact with each other; this is what we mean by ``quantum space-time''
and ``quantum geometry'' (Section~\ref{secqg}).
Finally, in Section~\ref{secoutlook}
a few more structures of a causal fermion system are
worked out which clarify the connection to other
approaches to non-regular space-time geometry
(Lorentzian length spaces, causal sets, lattices)
or might be useful for the geometric understanding and the future analysis of causal fermion systems.

\section{From Lorentzian Geometry to Causal Fermion Systems} \label{sectoCFS}
A basic concept behind causal fermion systems is to encode the geometry
in a measure on linear operators on a Hilbert space.
We now explain how to get into this setting starting from a usual four-dimensional space-time
(for other dimensions see~\cite[Section~1.4]{drum}, \cite[Section~4]{finite} and~\cite{topology}).
More precisely, let~$(\scrM, g)$ be a smooth, globally hyperbolic, time-oriented
Lorentzian spin manifold of dimension four.
For the signature of the metric we use the convention~$(+ ,-, -, -)$.
We denote the corresponding spinor bundle by~$S\scrM$. Its fibers~$S_p\scrM$ are endowed
with an inner product~$\Sl .|. \Sr_p$ of signature~$(2,2)$.
Clifford multiplication is described by a mapping~$\gamma$
which satisfies the anti-commutation relations,
\beq \label{cliffdef}
\gamma \::\: T_p\scrM \rightarrow \Lin(S_p\scrM) \qquad
\text{with} \qquad \gamma(u) \,\gamma(v) + \gamma(v) \,\gamma(u) = 2 \, g(u,v)\,\1_{S_p(\scrM)} \:.
\eeq
We also write Clifford multiplication in components with the Dirac matrices~$\gamma^j$.
The metric connections on the tangent bundle and the spinor bundle are denoted by~$\nabla$.
The sections of the spinor bundle are also referred to as wave functions.

We denote the $k$-times continuously differentiable sections of the spinor bundle by~$C^k(\scrM, S\scrM)$.
The Dirac operator~$\Dir$ is defined by
\[ \Dir := i \gamma^j \nabla_j \::\: C^\infty(\scrM, S\scrM) \rightarrow C^\infty(\scrM, S\scrM)\:. \]
Given a real parameter~$m \in \R$ (the ``mass''), the Dirac equation reads
\[ (\Dir - m) \,\psi = 0 \:. \]
We mainly consider solutions in the class~$\Cisc(\scrM, S\scrM)$ of smooth sections
with spatially compact support. On such solutions, one has the scalar product
\[ %\label{print}
(\psi | \phi)_m = 2 \pi \int_\scrN \Sl \psi \,|\, \gamma(\nu)\, \phi \Sr_p\: d\mu_\scrN(p) \:, \]
where~$\scrN$ denotes any Cauchy surface and~$\nu$ its future-directed normal
(due to current conservation, the scalar product is
in fact independent of the choice of~$\scrN$; for details see~\cite[Section~2]{finite}).
Forming the completion gives the Hilbert space~$(\H_m, (.|.)_m)$.

Next, we choose a closed subspace~$\H \subset \H_m$
of the solution space of the Dirac equation.
The induced scalar product on~$\H$ is denoted by~$\la .|. \ra_\H$.
There is the technical difficulty that the wave functions in~$\H$ are in general not continuous,
making it impossible to evaluate them pointwise.
For this reason, we need to introduce an {\em{ultraviolet regularization}} on
the length scale~$\varepsilon$, described mathematically by a linear
\[ \text{\em{regularization operator}} \qquad {\mathfrak{R}}_\varepsilon \::\: \H \rightarrow C^0(\scrM, S\scrM) \:. \]
In the simplest case, the regularization can be realized by a convolution
on a Cauchy surface or in space-time (for details see~\cite[Section~4]{finite}
or~\cite[Section~\S1.1.2]{cfs}). For us, the regularization is not just a technical tool,
but it realizes the concept already mentioned
at the beginning of this paper that we want to change the geometric structures on the microscopic
scale. With this in mind, we always consider the regularized quantities as those having mathematical and
physical significance. Different choices of regularization operators realize different
microscopic space-time structures.

Given~${\mathfrak{R}}_\varepsilon$, for any space-time point~$p \in \scrM$ we consider the bilinear form
\[ b_p \::\: \H \times \H \rightarrow \C\:,\quad
b_p(\psi, \phi) = -\Sl ({\mathfrak{R}}_\varepsilon\psi)(p) | ({\mathfrak{R}}_\varepsilon \phi)(p) \Sr_p \:. \]
This bilinear form is well-defined and bounded because~${\mathfrak{R}}_\varepsilon$ 
maps to the continuous wave functions and because evaluation at~$p$ gives a linear operator of finite rank.
Thus for any~$\phi \in \H$, the anti-linear form~$b_p(.,\phi) : \H \rightarrow \C$
is continuous. By the Fr{\'e}chet-Riesz theorem,
there is a unique~$\chi \in \H$
such that~$b_p(\psi,\phi) = \la \psi | \chi \ra_\H$ for all~$\psi \in \H$.
The mapping~$\phi \mapsto \chi$ is linear and bounded. We thus obtain a unique bounded linear
operator~$F^\varepsilon(p)$ on~$\H$ which is characterized by the relation
\beq \label{Fepsdef}
(\psi \,|\, F^\varepsilon(p)\, \phi) = -\Sl ({\mathfrak{R}}_\varepsilon\psi)(p) | 
({\mathfrak{R}}_\varepsilon \phi)(p) \Sr_p \qquad \text{for all~$\psi, \phi
\in \H$} \:.
\eeq
Taking into account that the inner product on the Dirac spinors at~$p$ has signature~$(2,2)$,
the local correlation operator~$F^\varepsilon(p)$ is a symmetric operator on~$\H$
of rank at most four, which (counting multiplicities) has at most two positive and at most two negative eigenvalues.
Varying the space-time point, we obtain a mapping
\[ F^\varepsilon \::\: \scrM \rightarrow \F \subset \Lin(\H)\:, \]
where~$\F$ denotes all symmetric operators of
rank at most four with at most two positive and at most two negative eigenvalues.
Finally, we introduce the
\[ \text{\em{universal measure}} \qquad d\rho := (F^\varepsilon)_* \,d\mu_\scrM \]
as the push-forward of the volume measure on~$\scrM$ under the mapping~$F^\varepsilon$
(thus~$\rho(\Omega) := \mu_\scrM((F^\varepsilon)^{-1}(\Omega))$).

In this way, we obtain a measure~$\rho$ on the set~$\F \subset \Lin(\H)$ of
linear operators on a Hilbert space~$\H$.
The basic concept is to work exclusively with these objects, but to drop all
other structures (like the Lorentzian metric~$g$, the structure of the spinor bundle~$S\scrM$,
the manifold structure of~$\scrM$, and even the structure of~$\scrM$ being a point set).
This leads us to the structure of a causal fermion system of spin dimension two,
as will be defined abstractly at the beginning of the next section.
%(see Definition~\ref{defparticle}).

Before turning attention to the abstract setting, we make a few comments
on the underlying physical picture.
The vectors in the subspace~$\H \subset \H_m$ have the interpretation
as those Dirac wave functions which are realized in the physical system under
consideration.
If we describe for example a system of one electron,
then the wave function of the electron is contained in~$\H$. Moreover, $\H$ includes all the wave functions
which form the so-called Dirac sea (for an explanation of this point
see for example~\cite{srev}). We refer to the vectors in~$\H$ as the
{\em{physical wave functions}}. The name causal {\em{fermion}} system
is motivated by the fact that Dirac particles are fermions.
According to~\eqref{Fepsdef}, 
the local correlation operator~$F^\varepsilon(p)$ describes 
densities and correlations of the physical
wave functions at the space-time point~$p$.
Working exclusively with the local correlation operators and the
corresponding push-forward measure~$\rho$ means in particular
that the geometric structures are encoded in and must be retrieved from the physical wave functions.
Since the physical wave functions describe the distribution of
matter in space-time, one can summarize this concept
by saying that {\em{matter encodes geometry}}.

\section{The Abstract Setting and a Few Inherent Structures} \label{secabstract}
\subsection{Basic Definition of a Causal Fermion System}

We now give the abstract definition of a causal fermion system:
\begin{Def} \label{defparticle}  {\em{ %\hspace*{1em}
Given a separable complex Hilbert space~$\H$ with scalar product~$\la .|. \ra_\H$
and a parameter~$n \in \N$ (the {\em{``spin dimension''}}), we let~$\F \subset \Lin(\H)$ be the set of all
self-adjoint operators on~$\H$ of finite rank, which (counting multiplicities) have
at most~$n$ positive and at most~$n$ negative eigenvalues. On~$\F$ we are given
a measure~$\rho$ (defined on a $\sigma$-algebra of subsets of~$\F$), the so-called
{\em{universal measure}}. We refer to~$(\H, \F, \rho)$ as a {\em{causal fermion system}}.
}}
\end{Def}

This definition, which was first given in~\cite[Section~1.2]{rrev},
no longer involves the usual geometric structures (like a manifold~$\scrM$, a Lorentzian metric~$g$
or a spinor bundle~$S\scrM$). Therefore, it is far from obvious why and how a causal fermion
system should provide a setting for geometry in space-time.
All we can say right away is that a causal fermion system encodes a lot of information.
For example, parts of this information can be retrieved by taking traces of products of
operators in~$\F$ and integrating,
\[ \int_\F \tr(x)\: d\rho(x)\:,\qquad \int_\F d\rho(x) \int_\F d\rho(y) \: \tr(x \!\cdot\! y) \:, \qquad \ldots \:. \]
The point is that by describing all this information with useful and convenient notions, one recovers
a space-time as well as geometric structures therein.
All these structures are {\em{inherent}} in the sense that they are no additional input,
but they only give information which is already encoded in the causal fermion system
an apposite name, thereby providing a better intuitive understanding of the causal fermion system.

Working with the inherent structures, causal fermion systems provide a general mathematical framework
in which there are many analytic, geometric and topological structures.
In particular, it becomes possible to generalize notions of differential geometry to the non-smooth setting.
From the physical point of view, causal fermion systems
are a proposal for quantum geometry and an approach to quantum gravity.
Giving quantum mechanics, general relativity and quantum field theory as limiting cases, they are
a candidate for a unified physical theory.
For the physical applications, the key point is that the physical equations can be formulated
within the setting of causal fermion systems
in terms of a variational principle called the {\em{causal action principle}}
(see~\cite[Section~1.1]{cfs}).
The causal action principle is the analytic core of the theory.
For brevity we cannot introduce the causal action principle here,
but instead refer the interested reader to the 
mathematical introduction in the recent paper~\cite{jet} and to the references therein.

\subsection{Space-Time and Causal Structure}
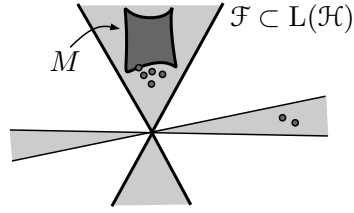
\begin{figure} % figcfs.svg
% \usepackage[usenames,dvipsnames]{pstricks}
% \usepackage{epsfig}
% \usepackage{pst-grad} % For gradients
% \usepackage{pst-plot} % For axes
% \usepackage[space]{grffile} % For spaces in paths
% \usepackage{etoolbox} % For spaces in paths
% \makeatletter % For spaces in paths
% \patchcmd\Gread@eps{\@inputcheck#1 }{\@inputcheck"#1"\relax}{}{}
% \makeatother
% 
\psscalebox{1.0 1.0} % Change this value to rescale the drawing.
{
\begin{pspicture}(0,-1.3610485)(3.6917334,1.3610485) %(0,-1.3610485)(6.6917334,1.3610485)
\definecolor{colour0}{rgb}{0.8,0.8,0.8}
\definecolor{colour1}{rgb}{0.4,0.4,0.4}
\pspolygon[linecolor=white, linewidth=0.02, fillstyle=solid,fillcolor=colour0](0.91751134,1.3317788)(1.9034183,-0.41748416)(2.8348446,1.332738)
\psbezier[linecolor=black, linewidth=0.04, fillstyle=solid,fillcolor=colour1](1.553067,0.89983374)(1.5447986,1.547851)(1.4361526,1.1536276)(1.8759718,1.1576269552442744)(2.315791,1.1616263)(2.0934794,1.5804809)(2.1575115,0.90066475)(2.2215433,0.22084858)(2.306556,0.64546037)(1.8751786,0.54429364)(1.4438012,0.4431269)(1.5613352,0.25181648)(1.553067,0.89983374)
\rput[bl](0.52728915,0.4549603){$M$}
\pscircle[linecolor=black, linewidth=0.02, fillstyle=solid,fillcolor=colour1, dimen=outer](1.7152891,0.4931825){0.05111111}
\pscircle[linecolor=black, linewidth=0.02, fillstyle=solid,fillcolor=colour1, dimen=outer](1.7819558,0.39096028){0.05111111}
\pscircle[linecolor=black, linewidth=0.02, fillstyle=solid,fillcolor=colour1, dimen=outer](1.8930669,0.43984917){0.05111111}
\pscircle[linecolor=black, linewidth=0.02, fillstyle=solid,fillcolor=colour1, dimen=outer](1.884178,0.27540472){0.05111111}
\pscircle[linecolor=black, linewidth=0.02, fillstyle=solid,fillcolor=colour1, dimen=outer](2.0308447,0.40429363){0.05111111}
\pspolygon[linecolor=white, linewidth=0.02, fillstyle=solid,fillcolor=colour0](2.448178,-1.3365248)(1.8933822,-0.34726194)(1.3664002,-1.3508174)
\psline[linecolor=black, linewidth=0.04](0.904178,1.3406092)(1.892704,-0.36415082)(2.8375113,1.3514048)
\pspolygon[linecolor=colour0, linewidth=0.02, fillstyle=solid,fillcolor=colour0](1.953067,-0.35029832)(4.197794,0.09984918)(4.224178,-0.4068175)
\pspolygon[linecolor=colour0, linewidth=0.02, fillstyle=solid,fillcolor=colour0](0.010844693,-0.3490397)(1.8375113,-0.37570637)(0.041955803,-0.7357064)
\psline[linecolor=black, linewidth=0.02](0.0019558037,-0.33570638)(4.2286224,-0.41570637)
\psline[linecolor=black, linewidth=0.02](0.02862247,-0.7490397)(4.197511,0.11318251)
\psline[linecolor=black, linewidth=0.04](2.4064002,-1.331133)(1.8934299,-0.37748414)(1.3575114,-1.3508174)
\pscircle[linecolor=black, linewidth=0.02, fillstyle=solid,fillcolor=colour1, dimen=outer](3.6264002,-0.17348416){0.05111111}
\pscircle[linecolor=black, linewidth=0.02, fillstyle=solid,fillcolor=colour1, dimen=outer](3.795289,-0.24015082){0.05111111}
\rput[bl](2.9317336,0.9438492){$\F \subset \Lin(\H)$}
\psbezier[linecolor=black, linewidth=0.02, arrowsize=0.05291667cm 2.0,arrowlength=1.4,arrowinset=0.0]{<-}(1.4669569,1.0098974)(1.1542009,1.0733252)(1.0045664,0.96429986)(0.8036214,0.7631343899628382)
\end{pspicture}
}
\caption{Space-time~$M$ of a causal fermion system}
\label{figcfs}
\end{figure}

We now introduce a few inherent structures of a causal fermion system~$(\H, \F, \rho)$.
{\em{Space-time}}~$M$ is defined as the support of the universal
measure\footnote{The {\em{support}} of a measure is defined as the complement of the largest open
set of measure zero, i.e.
\[ \supp \rho := \F \setminus \bigcup \big\{ \text{$\Omega \subset \F$ \,\big|\,
$\Omega$ is open and $\rho(\Omega)=0$} \big\} \:. \]
It is by definition a closed set.},
\beq \label{Mdef}
M := \text{supp}\, \rho \subset \F \:.
\eeq
On~$M$ we consider the topology induced by~$\F$ (generated by the $\sup$-norm
on~$\Lin(\H)$).
Typically, $M$ is a low-dimensional subset of~$\F$ which can
have smooth but also discrete or non-regular components
(see Figure~\ref{figcfs}).

The next definition is the reason for the name {\em{causal}} fermion system.
\begin{Def} (causal structure) \label{def2}
{\em{ For any~$x, y \in M$, the product~$x y$ is an operator
of rank at most~$2n$. We denote its non-trivial eigenvalues (counting algebraic multi\-pli\-ci\-ties)
by~$\lambda^{xy}_1, \ldots, \lambda^{xy}_{2n}$.
The points~$x$ and~$y$ are
called {\em{spacelike}} separated if all the~$\lambda^{xy}_j$ have the same absolute value.
They are said to be {\em{timelike}} separated if the~$\lambda^{xy}_j$ are all real and do not all 
have the same absolute value.
In all other cases (i.e.\ if the~$\lambda^{xy}_j$ are not all real and do not all 
have the same absolute value),
the points~$x$ and~$y$ are said to be {\em{lightlike}} separated. }}
\end{Def} \noindent
We remark that this definition is closely linked to the structure of the causal action
(see~\cite[\S1.1.2]{cfs}).

A causal fermion system also distinguishes a {\em{direction of time}}.
To this end, we let~$\pi_x$ be the orthogonal projection in~$\H$ on the subspace~$x(\H) \subset \H$
and introduce the functional
\beq \label{Cform}
{\mathscr{C}} \::\: M \times M \rightarrow \R\:,\qquad
{\mathscr{C}}(x, y) := i \tr \big( y\,x \,\pi_y\, \pi_x - x\,y\,\pi_x \,\pi_y \big)
\eeq
(this functional was first stated in~\cite[Section~8.5]{topology}, motivated by constructions
in~\cite[Section~3.5]{lqg}).
Obviously, this functional is anti-symmetric in its two arguments.
This makes it possible to introduce the notions
\beq \label{tdir}
\left\{ \begin{array}{cl} \text{$y$ lies in the {\em{future}} of~$x$} &\quad \text{if~${\mathscr{C}}(x, y)>0$} \\[0.2em]
\text{$y$ lies in the {\em{past}} of~$x$} &\quad \text{if~${\mathscr{C}}(x, y)<0$}\:. \end{array} \right.
\eeq

We remark that in suitable limiting cases, the above notions of causality indeed agree with the
causal structure of a Lorentzian space-time (for details see~\cite[\S1.2.5]{cfs} or~\cite[Sections~4 and~5]{lqg}).

\subsection{Spinors and Physical Wave Functions}
For every~$x \in \F$ we define the {\em{spin space}}~$S_xM$ by~$S_xM = x(\H)$; it is a subspace of~$\H$ of
dimension at most~$2n$. On the spin space~$S_xM$ the {\em{spin scalar product}} $\Sl .|. \Sr_x$
is defined by
\beq \label{ssp}
\Sl . | . \Sr_x : S_xM \times S_xM \rightarrow \C \:,\qquad
\Sl u | v \Sr_x = -\la u | x u \ra_\H\:,
\eeq
making the spin space~$(S_xM, \Sl .|. \Sr_x)$ to an indefinite inner product of
signature~$(p,q)$ with~$p,q \leq n$.

A {\em{wave function}}~$\psi$ is defined as a function which to every~$x \in M$ associates a vector of the
corresponding spin space,
\beq \label{psirep}
\psi \::\: M \rightarrow \H \qquad \text{with} \qquad \psi(x) \in S_xM \quad \text{for all~$x \in M$}\:.
\eeq
To every vector~$u \in \H$ we can associate a wave function~$\psi^u$
obtained by taking the orthogonal projections to the corresponding spin spaces,
\[ \psi^u \::\: M \rightarrow \H\:,\qquad \psi^u(x) = \pi_x u \in S_xM \]
(with~$\pi_x$ as defined before~\eqref{Cform}).
We refer to~$\psi^u$ as the {\em{physical wave function}} of~$u \in \H$
(we remark that, for the causal fermion systems obtained by the construction
in Section~\ref{sectoCFS}, after suitable identifications made,
the physical wave function~$\psi^u$ indeed coincides with
the regularized Dirac wave function~${\mathfrak{R}}_\varepsilon u$;
for details see~\cite[\S1.2.4]{cfs}).

In this way, every vector of~$\H$ can be represented by a wave function in space-time.
The resulting ensemble of all physical wave functions is described most conveniently
by the {\em{kernel of the fermionic projector}}~$P(x,y)$ defined by
\beq \label{Pxydef}
P(x,y) = \pi_x \,y|_{S_yM} \::\: S_yM \rightarrow S_xM\:.
\eeq
Due to the factor~$x$ in the definition of the spin scalar product~\eqref{ssp}, the kernel of the fermionic
projector is symmetric in the sense that $P(x,y)^* = P(y,x)$, as is immediately verified by the computation
\begin{align*}
\Sl u \,|\, P(x,y) \,v \Sr_x &= - \la u \,|\, x\, P(x,y) \,v \ra_\H = - \la u \,|\, x y \,v \ra_\H \\
&= -\la \pi_y \,x\, u \,|\, y \,v \ra_\H = \Sl P(y,x)\, u \,|\,  v \Sr_y \qquad (\text{for~$u \in S_xM$, $v \in S_yM$)}\:.
\end{align*}

\section{Geometric Structures of a Causal Fermion System} \label{seclqg}
In~\cite{lqg} the geometric structures of a causal fermion system 
were worked out abstractly, and it was shown that in a certain limiting case one recovers
Lorentzian spin geometry. We now outline a few constructions and results from this paper.

\subsection{Regular Causal Fermion Systems}
According to Definition~\ref{defparticle}, the operators in~$M \subset \F$ have at most~$n$
positive and at most~$n$ negative eigenvalues. In most situations of physical interest
and most examples, the number of positive and negative eigenvalues equals~$n$.
This motivates the following definition:
\begin{Def} \label{defregular} {\em{
A space-time point~$x \in M$ is said to be {\em{regular}} if~$x$ has the maximal possible rank,
i.e.~$\dim x(\H) = 2n$. Otherwise, the space-time point is called {\em{singular}}.
A causal fermion system is {\em{regular}} if all its space-time points are regular.}}
\end{Def} \noindent
For a regular causal fermion system, the set
\[ SM := \bigcup_{x \in M} S_xM \]
has the structure of a topological vector bundle with base~$M$ and fibers~$S_xM$
(for details and all topological issues see~\cite{topology}).
The fibers are endowed with an inner product~$\Sl .|. \Sr_p$ of signature~$(n,n)$.
A wave function~\eqref{psirep} is a section of this bundle
(for the notion of continuity of such wave functions and the space~$C^0(M, SM)$
we refer to~\cite[\S1.1.4]{cfs}).
As a specific feature, all the fibers are subspaces of the same
Hilbert space~$(\H, \la .|. \ra_\H$).

\subsection{Clifford Subspaces} \label{seccliff}
The structure of wave functions~\eqref{psirep} taking values in the spin spaces
is reminiscent of sections of a vector bundle. 
However, one important structure is missing: we have no Dirac matrices
and no notion of Clifford multiplication. The following definition is a step towards
introducing these additional structures.

\begin{Def} (Clifford subspace) \label{defcliffsubspace} {\em{
We denote the space of symmetric linear operators on~$(S_xM, \Sl .|. \Sr_x)$
by~$\Symm(S_xM) \subset \Lin(S_xM)$.
A subspace~$K \subset \Symm(S_xM)$ is called
a {\em{Clifford subspace}} of signature~$(r,s)$ at the point~$x$ (with~$r,s \in \N_0$)
if the following conditions hold:
\begin{itemize}[leftmargin=2.5em]
\item[(i)] For any~$u, v \in K$, the anti-commutator~$\{ u,v \} \equiv u v + v u$ is a multiple
of the identity on~$S_xM$.
\item[(ii)] The bilinear form~$\la .,. \ra$ on~$K$ defined by
\beq \label{anticomm}
\frac{1}{2} \left\{ u,v \right\} = \la u,v \ra \, \1 \qquad {\text{for all~$u,v \in K$}}
\eeq
is non-degenerate and has signature~$(r,s)$.
\end{itemize} }}
\end{Def}
In view of the anti-commutation relations~\eqref{anticomm}, a Clifford subspace can be
regarded as a generalization of the space spanned by the usual Dirac matrices.
However, the above definition has two shortcomings:
First, there are many different Clifford subspaces, so that there is no unique
notion of Clifford multiplication. Second, we are missing the structure of tangent vectors.
We now explain how to overcome these shortcomings.

\subsection{Connection and Curvature} \label{secconn}
The kernel of the kernel of the fermionic projector~\eqref{Pxydef} is a mapping from
one spin space to another and thus induces relations between space-time points.
The idea is to use these relations for the construction of a spin connection~$D_{x,y}$, being a unitary
mapping between the corresponding spin spaces,
\[ D_{x,y} \::\: S_yM \rightarrow S_xM \]
(we consistently use the notation that
the subscript~$_{xy}$ denotes an object at the point~$x$, whereas
the additional comma $_{x,y}$ denotes an operator which maps an object at~$y$
to an object at~$x$). The simplest idea for the construction of the spin connection 
would be to form a polar decomposition~$P(x,y) = S\,U$ with a symmetric operator~$S : S_xM \rightarrow S_xM$
and a unitary operator~$U : S_yM \rightarrow S_xM$ and to
introduce the spin connection as the unitary part, $D_{x,y} = U$.
However, this method is too naive, because we want the spin connection to be compatible
with a corresponding metric connection~$\nabla_{x,y}$ which should map Clifford subspaces
at~$x$ and~$y$ (see Definition~\ref{defcliffsubspace} above) isometrically to each other.
Another complication is that, as explained after Definition~\ref{defcliffsubspace},
the Clifford subspaces at~$x$ and~$y$ are not unique.
The method to bypass these problems is to work with several Clifford subspaces
and to use so-called splice maps, as we now briefly explain.

First, it is useful to restrict the freedom in choosing the Clifford subspaces
with the following construction.
Recall that for any~$x \in M$, the operator~$(-x)$ on~$\H$ has at most $n$~positive
and at most~$n$ negative eigenvalues. We denote its positive and negative spectral subspaces
by~$S_x^+M$ and~$S_x^-M$, respectively. In view of~\eqref{ssp}, these subspaces are also orthogonal
with respect to the spin scalar product,
\[ S_xM = S_x^+M \oplus S_x^-M \:. \]
We introduce the {\em{Euclidean sign operator}}~$s_x$ as a symmetric operator on~$S_xM$
whose eigenspaces corresponding to the eigenvalues~$\pm 1$ are the spaces~$S_x^+M$
and~$S_x^-M$, respectively.
Since~$s_x^2=\1$, the span of the Euclidean sign operator is a one-dimensional Clifford subspace of
signature~$(1,0$). The idea is to extend~$s_x$ to obtain higher-dimensional Clifford subspaces.
We thus define a {\em{Clifford extension}} as a Clifford subspace which contains~$s_x$.
By restricting attention to Clifford extensions, we have reduced the freedom in choosing
Clifford subspaces. However, still there is not a unique Clifford extension, even for fixed
dimension and signature. But one can define the {\em{tangent space}}~$T_x$
as an equivalence class of Clifford extensions; for details see~\cite[Section~3.1]{lqg}.
Choosing~$r=1$, the bilinear form~$\la .,. \ra$ in~\eqref{anticomm} induces a Lorentzian metric on the
tangent space.

In the remainder of this section, we assume that the causal fermion system is regular
(see Definition~\ref{defregular}). Moreover, we need the following stronger version
of timelike separation:
\begin{Def} \label{defproptl} {\em{
The space-time point~$x \in M$ is {\em{properly timelike}}
separated from~$y \in M$ if the closed chain~$A_{xy}$ defined by
\[ %\label{Axydef}
A_{xy} = P(x,y)\, P(y,x) \::\: S_xM \rightarrow S_xM \]
has a strictly positive spectrum and if all eigenspaces are (either positive or negative) definite subspaces of~$(S_xM, \Sl .|. \Sr_x)$. }}
\end{Def} \noindent
The two following observations explain why the last definition makes sense
(for the proof see~\cite[\S1.1.6]{cfs}):
\begin{itemize}[leftmargin=2em]
\itemD Properly timelike separation implies timelike separation (see Definition~\ref{def2}).
\itemD The notion is symmetric in~$x$ and~$y$.
\end{itemize}

So far, the construction of the spin connection has been worked out only in the case
of spin dimension~$n=2$. Then for two properly timelike separated points~$x,y \in M$,
the spin space~$S_xM$ can be decomposed uniquely
into an orthogonal direct sum~$S_xM = I^+ \oplus I^-$ of a two-dimensional
positive definite subspace~$I^+$ and a two-dimensional negative definite subspace~$I^-$ of~$A_{xy}$.
We define the {\em{directional sign operator}}~$v_{xy}$ of~$A_{xy}$ as the
unique operator on~$S_xM$ such that
the eigenspaces corresponding to the eigenvalues~$\pm 1$
are the subspaces~$I^\pm$.

Having the Euclidean sign operator~$s_x$ and the directional sign operator~$v_{xy}$
to our disposal, under generic assumptions one can distinguish two Clifford subspaces at
the point~$x$: a Clifford subspace~$K_{xy}$ containing~$v_{xy}$ and a Clifford extension~$K_x^{(y)}$
(for details see~\cite[Lemma~3.12]{lqg}). Similarly, at the point~$y$ we have a
distinguished Clifford subspace~$K_{yx}$ (which contains~$v_{yx}$)
and a distinguished Clifford extension~$K_y^{(x)}$.
For the construction of the {\em{spin connection}}~$D_{x,y} : S_yM \rightarrow S_xM$ one works
with the Clifford subspaces~$K_{xy}$ and~$K_{yx}$ and demands that these are mapped
to each other. More precisely, the spin connection is uniquely characterized by the following
properties (see~\cite[Theorem~3.20]{lqg}):
\begin{itemize}[leftmargin=2em]
\item[(i)] $D_{x,y}$ is of the form
\[ D_{x,y} = e^{i \varphi_{xy}\, v_{xy}}\: A_{xy}^{-\frac{1}{2}}\: P(x,y) \quad \text{with} \quad
\varphi_{xy} \in \Big( -\frac{3\pi}{4}, -\frac{\pi}{2} \Big) \cup \Big( \frac{\pi}{2}, \frac{3\pi}{4} \Big) \:. \]
\item[(ii)] The spin connection maps the Clifford subspaces~$K_{xy}$ and~$K_{yx}$ to each other, i.e.\
\[ D_{y,x} \,K_{xy}\, D_{x,y} = K_{yx} \:. \]
\end{itemize}
The spin connection has the properties
\[ D_{y,x} = (D_{x,y})^{-1} = (D_{x,y})^* \qquad \text{and} \qquad
A_{xy} = D_{x,y}\, A_{yx}\, D_{y,x} \:. \]
All the assumptions needed for the construction of the spin connection are combined
in the notion that~$x$ and~$y$ must be {\em{spin-connectable}}
(see~\cite[Definition~3.17]{lqg}).

By composing the spin connection along a ``discrete path'' of space-time points,
one obtains a ``parallel transport'' of spinors.
When doing so, it is important to keep track of the different Clifford subspaces and to carefully transform
them to each other. In order to illustrate in an example how this works,
suppose that we want to compose the spin connection~$D_{y,z}$
with~$D_{z,x}$. As mentioned above, the spin connection~$D_{z,x}$ at the point~$z$
is constructed using the Clifford subspace~$K_{zx}$. The spin connection~$D_{y,z}$, however,
takes at the same space-time point~$z$ the Clifford subspace~$K_{zy}$ as reference.
This entails that before applying~$D_{y,z}$ we must transform
from the Clifford subspace~$K_{zx}$ to the Clifford subspace~$K_{zy}$.
This is accomplished by the {\em{splice map}}~$U_z^{(y|x)}$,
being a uniquely defined unitary transformation of~$S_xM$ with the property that
\[ K_{zy} = U_z^{(y|x)} \,K_{zx}\, \big( U_z^{(y|x)} \big)^* \:. \]
The splice map must be sandwiched between the spin connections in combinations like
\[ D_{y,z}\, U_z^{(y|x)} \,D_{z,x} \:. \]

In order to construct a corresponding metric connection~$\nabla_{x,y}$,
one uses a similar procedure to
relate the Clifford subspaces to corresponding Clifford extensions. More precisely, one
first unitarily transforms the Clifford extension~$K_y^{(x)}$ to the Clifford subspace~$K_{yx}$.
Unitarily transforming with the spin connection~$D_{xy}$ gives the Clifford subspace~$K_{xy}$.
Finally, one unitarily transforms to the Clifford extension~$K_x^{(y)}$.
Since the Clifford extensions at the beginning and end are
representatives of the corresponding tangent spaces, we thus obtain an isometry
\[ \nabla_{x,y} \::\: T_y \rightarrow T_x \]
between the tangent spaces (for details see~\cite[Section~3.4]{lqg}).

In this setting, {\em{curvature}} is defined as usual as the holonomy of the connection.
Thus the curvature of the spin connection is given by
\[ \mathfrak{R}(x,y,z) = U_x^{(z|y)} \:D_{x,y}\: U_y^{(x|z)} \:D_{y,z}\: U_z^{(y|x)}
\:D_{z,x} \::\: S_xM \rightarrow S_xM \:, \]
and similarly for the metric connection.

\subsection{Correspondence to Lorentzian Geometry} \label{seccorr}
In~\cite[Sections~4 and~5]{lqg} it is proven that the above notions
reduce to the metric connections and the Riemannian curvature on a globally hyperbolic Lorentzian
manifold if the causal fermion system is constructed by regularizing solutions of the Dirac equation
(similar as explained in Section~\ref{sectoCFS}) and removing the regularization
in a suitable way. These results show that the notions of connection and curvature
defined above indeed generalize the corresponding notions in Lorentzian spin geometry.

As in Section~\ref{sectoCFS}, we begin with a globally hyperbolic space-time~$(\scrM, g)$.
In order to keep the setting as simple as possible, in~\cite{lqg} it was assumed that~$(\scrM,g)$
is isometric to Minkowski space in the past of a Cauchy-hypersurface $\nn$.
This has the technical advantage that one can
work with simple and explicit regularization operators~${\mathcal{R}}_\varepsilon$
defined in Minkowski space
(more precisely, one works with $i \varepsilon$-regularization of a Dirac see structure;
see also~\cite[\S2.4.1]{cfs}).
Due to the ultraviolet regularization, the causal fermion system only describes
the ``coarse geometry'' down to the length scale~$\varepsilon$.
Therefore, in order to recover Lorentzian spin geometry, we need to analyze the
limit~$\varepsilon \searrow 0$.

For sufficiently small~$\varepsilon$, one can identify objects of Lorentzian spin geometry
with corresponding objects of the causal fermion system in the following way: \\
\begin{center}
\begin{tabular}{|c|c|}
\hline & \\[-0.8em]
$\quad$ {\bf{Lorentzian geometry}} $\quad$ & {\bf{causal fermion system}} \\[0.2em]
\hline & \\[-0.8em]
space-time manifold~$\scrM$ & space-time $M :=\supp \rho \subset \F \subset \Lin(\H)$ \\[0.2em]
space-time point~$p \in \scrM$ & space-time point~$x:=F^\varepsilon(p) \in M$ \\[0.2em]
spinor space $S_p\scrM$ & spin space $S_xM :=x(\H) \subset \H$ \\[0.2em]
wave function~$u \in \H \subset \H_m$ & physical wave function~$\psi^u : M \rightarrow \H$ \\
spinor~$u(p) \in S_p\scrM$ & spinor $\psi^u(x) \in S_xM$ \\[0.2em]
%topology of~$\scrM$ & topology of~$M^\varepsilon$ \\[0.2em]
%causal structure of Minkowski space & causal structure of Definition~\ref{def2} \\[0.2em]
\hline
\end{tabular}
\end{center}

\vspace*{0.5em}

\noindent
Assume that~$\gamma : [0,T] \rightarrow \scrM$
is a future-directed timelike curve
joining the points~$q:=\gamma(0)$ with~$p:=\gamma(T)$, for simplicity parametrized by arc length.
In order to avoid confusion with the different connections, we now denote the metric connection
on the tangent bundle~$T\scrM$ (i.e.\ the Levi-Civita connection) by~$\nablaLC$.
Likewise, the metric connection on the spinor bundle~$S \scrM$ is denoted by~$\DLC$.

In order to compare the parallel transports~$\DLC_{p,q}$ and $\nablaLC$ 
along~$\gamma$ with the corresponding connections~$D$ and~$\nabla$ of the 
causal fermion system, given~$N \in \N$ we choose intermediate points on the curve by
\beq \label{pndef}
p_n := \gamma \Big( \frac{n T}{N}\Big) \:,
\qquad n=0, \ldots, N\:.
\eeq
The following results show that in the limit~$\varepsilon \searrow 0$
and~$N \rightarrow \infty$, the connections agree, up to higher orders in the curvature tensor
(for the proof see~\cite[Theorem~5.12 and Corollary~5.13]{lqg}).

\begin{Thm} \label{thmgrav}
There is a subset of curves which is dense in the~$C^\infty$-topology
with the following property: Choosing~$N \in \N$ sufficiently large and~$\varepsilon>0$ 
sufficiently small, the points~$p_{n}$ and~$p_{n-1}$ are spin-connectable for all~$n=1,\ldots, N$, 
and every point $p_n$ lies in the future of $p_{n-1}$ (in the sense~\eqref{tdir}). Moreover,
\begin{align*}
\lim_{N \rightarrow \infty} &\:\lim_{\varepsilon \searrow 0} %D^{N,\varepsilon}_{(p,q)}
D_{(p_N, p_{N-1})} \:D_{(p_{N-1}, p_{N-2})} \:\cdots\: D_{(p_1, p_0)} \\
& =  \DLC_{p,q} \, \Texp \bigg( \frac{1}{6} \int_\gamma \Big(m^2 - \frac{\s}{12} \Big)^{-1} \\
&\qquad \times \DLC_{q,\gamma(t)}
\Big[\epsilon_j\, (\nabla_{e_j} R)\big(\dot{\gamma}(t), e_j \big)\, \dot{\gamma}(t) \Big]
\cliff \,\dot{\gamma}(t) \cliff \DLC_{\gamma(t), q} \: dt \bigg) \\
\lim_{N \rightarrow \infty} &\:\lim_{\varepsilon \searrow 0}
\nabla_{p_N, p_{N-1}} \:\nabla_{p_{N-1}, p_{N-2}}
\:\cdots\: \nabla_{p_1, p_0} \\
&=  \nablaLC_{q,p} + \O \!\left( L(\gamma)\: \frac{\| \nabla R\|}{m^2} \right)
\Big(1 + \O \Big( \frac{\s}{m^2} \Big) \Big) \,,
\end{align*}
where~$\Texp$ is the time-ordered exponential, and~$L(\gamma)$ is the length of the curve~$\gamma$.
\end{Thm}

\section{Going Beyond Lorentzian Geometry} \label{secbeyond}
\subsection{Non-Smooth Geometries} \label{secnoreg}
The construction of connection and curvature in Section~\ref{secconn} did no rely
on a smooth space-time structure. Indeed, these constructions even apply in
non-regular or discrete space-times (for many examples see~\cite[Section~9]{topology}).
Therefore, causal fermion systems are a framework for non-smooth geometries.
We now explain in words how this works and what to keep in mind.
Suppose that a causal fermion system~$(\H, \F, \rho)$ is given.
If space-time~$M:= \text{supp}\, \rho$ has the structure of a smooth manifold,
then we are in the smooth setting.
Otherwise, $M$ merely is a topological space, with additional structures
induced by the fact that the space-time points are linear operators in~$\F$.
The singular points (see Definition~\ref{defregular}) must be treated separately.
Removing them from~$M$ (by multiplying~$\rho$ with the characteristic function
of the regular points), one gets a regular causal fermion system.
Then the constructions in Sections~\ref{seccliff} and~\ref{secconn} apply,
giving a spin connection~$D_{x,y}$, a metric connection~$\nabla_{x,y}$
as well as corresponding curvatures.

A subtle point to keep in mind is that, since~$M$ has no manifold structure,
there is no notion of tangent vectors. This also means that there is
no analog of Clifford multiplication~\eqref{cliffdef}.
Instead, the Clifford subspace (see Definition~\ref{defcliffsubspace})
merely is a vector space of linear operators acting on the spin space
endowed with an inner product~$\la .,. \ra$, but the vectors in the
Clifford subspace are not related to usual tangent vectors
defined as equivalence classes of curves or as derivations.

One method for getting along without tangent vectors is to
work with {\em{tangent cone measures}}, as we now briefly outline
(another method, which will
be explained in Section~\ref{seclls} below, is to
construct a corresponding Lorentzian length space).
In the simplest version of this construction, one first chooses a mapping~${\mathcal{A}}$
from space-time to the vector space~$\Symm(S_x)$ of all symmetric linear operators on the spin space,
\beq \label{Afunct}
{\mathcal{A}} : M \rightarrow \Symm(S_x) \qquad \text{with} \qquad 
{\mathcal{A}}(u)= \pi_x \,(y-x)\, x |_{S_x}\:.
\eeq
Then the push-forward~${\mathcal{A}}_*(\rho)$ is a measure on~$\Symm(S_x)$.
A {\em{conical set}} is a subset of~$\Symm(S_x)$ with~$\R^+ A = A$.
The intuitive idea is to define the tangent cone measure~$\mu_\con$ at the space-time point~$x$
as a measure on conical sets obtained by restriction to smaller and smaller neighborhoods of~$x$
and rescaling, i.e.\ 
\[ \mu_\con(A) := \liminf_{\delta \searrow 0}\:
\frac{1}{\rho \big( B_\delta(x) \big)}\: 
\rho \Big( {\mathcal{A}}^{-1}(A_k) \cap B_\delta(x) \Big) \]
(where~$B_\delta(x) \subset \Lin(\H)$ is the Banach space ball).
In order to get $\sigma$-additivity, this definition must be modified
and complemented by measure-theoretic constructions (for details see~\cite[Section~6]{topology}).
Such tangent cone measures make it possible to analyze the local structure of space-time in a
neighborhood of a point~$x \in M$, again without any differentiability assumptions.
In particular, the tangent cone measures can be used to distinguish a specific Clifford subspace~$\Cl_x$
and to relate~$\Cl_x$ to neighboring space-time points (for details see again~\cite[Section~6]{topology}).

\subsection{Quantum Geometries} \label{secqg}
The general definition of causal fermion systems (see Definition~\ref{defparticle})
covers many situation in which space-time $M:=\supp \rho$ does not have the
structure of a manifold or a space-time lattice
and cannot be thought of as a classical space-time.
Since in our setting the geometric structure are encoded in the
quantum matter, we subsume the general situation under the notion {\em{quantum geometry}}.
This name is also motivated by the fact that these more general space-times are
indeed relevant for the applications to quantum field theory (see the mechanism of {\em{microscopic mixing}}
in~\cite{qft} or the related {\em{fragmentation of space-times}} in~\cite{perturb}).

We now illustrate in a simple example what a ``quantum space-time'' is about.
To this end, we generalize the setting of Section~\ref{sectoCFS} by considering on~$\scrM$
a family of Lorentzian metrics~$g_\tau$ indexed by a parameter~$\tau \in [0,1]$
(for example, the family can be obtained by varying the metric inside a compact subset~$K \subset \scrM$).
We assume that for every~$\tau$, the manifold~$(\scrM, g_\tau)$ 
is globally hyperbolic, time-oriented and spin. 
Moreover, we identify the Hilbert spaces~$\H_m$ for different values of~$\tau$
(for example by identifying boundary values on a Cauchy surface which does not intersect~$K$).
Then the construction of Section~\ref{sectoCFS} gives for every~$\tau \in [0,1]$
a measure~$\rho_\tau$ on~$\F$. The equation
\[ \tilde{\rho}(\Omega) = \int_0^1 \rho_\tau(\Omega)\: d\tau \]
(where~$d\tau$ is the Lebesgue measure) defines a measure on~$\F$.
The resulting space-time~$\tilde{M}$ as given by
\[ \tilde{M} := \supp \tilde{\rho} = \overline{\bigcup_{\tau \in (0,1)} M_\tau} \qquad \text{with} \qquad
M_\tau := \supp \rho_\tau \]
can no longer be identified with the original space-time manifold~$\scrM$,
but it can be thought of as a collection of the whole family of space-times~$(\scrM, g_\tau)_{\tau \in [0,1]}$. 
\begin{figure} % figqg.svg
% \usepackage[usenames,dvipsnames]{pstricks}
% \usepackage{epsfig}
% \usepackage{pst-grad} % For gradients
% \usepackage{pst-plot} % For axes
% \usepackage[space]{grffile} % For spaces in paths
% \usepackage{etoolbox} % For spaces in paths
% \makeatletter % For spaces in paths
% \patchcmd\Gread@eps{\@inputcheck#1 }{\@inputcheck"#1"\relax}{}{}
% \makeatother
% 
\psscalebox{1.0 1.0} % Change this value to rescale the drawing.
{
\begin{pspicture}(0,0.7)(4.207253,2.7503335) %(0,-2.7503335)(8.207253,2.7503335)
\definecolor{colour1}{rgb}{0.8,0.8,0.8}
\definecolor{colour4}{rgb}{0.4,0.4,0.4}
\definecolor{colour2}{rgb}{0.0,0.8,0.0}
\definecolor{colour3}{rgb}{0.0,0.8039216,0.0}
\pspolygon[linecolor=white, linewidth=0.02, fillstyle=solid,fillcolor=colour1](0.024142016,2.7385485)(2.2268088,0.5367779)(4.4125867,2.7403336)
\psline[linecolor=black, linewidth=0.04](0.015253127,2.720778)(2.2019198,0.5545557)(4.40192,2.727889)
\rput[bl](2.9228086,2.3292224){$\tilde{M}$}
\psframe[linecolor=white, linewidth=0.04, fillstyle=solid, dimen=outer](3.6885865,-0.6214443)(1.5596976,-2.7503333)
\pspolygon[linecolor=colour4, linewidth=0.02, fillstyle=solid,fillcolor=colour4](0.73525316,2.1207778)(0.8419198,2.2852223)(0.9752531,2.4096668)(1.0508087,2.4852223)(1.3619198,2.4896667)(1.6774753,2.480778)(2.064142,2.440778)(2.3263643,2.391889)(2.6196976,2.3385556)(2.9619198,2.2363334)(3.3263643,2.080778)(3.1974754,1.8985556)(3.055253,1.7163335)(2.8819199,1.8052224)(2.6819198,1.8541112)(2.4063642,1.8541112)(2.104142,1.8941113)(1.8419198,1.9874445)(1.5663643,2.0230002)(1.304142,2.031889)(1.0908086,2.0541112)
\psbezier[linecolor=colour4, linewidth=0.04](1.0108087,2.4452224)(1.3386452,2.486797)(1.6811622,2.492369)(2.1943076,2.418555688858037)(2.707453,2.3447425)(2.911377,2.2571213)(3.3219197,2.0630002)
\psbezier[linecolor=black, linewidth=0.02](0.8908087,2.3341112)(1.256924,2.3542848)(1.6412027,2.3447082)(2.0063643,2.2807779110802526)(2.3715258,2.2168477)(2.6995964,2.133228)(3.2019198,1.8941113)
\psbezier[linecolor=colour4, linewidth=0.04](0.7752531,2.151889)(1.2808362,1.9114366)(1.467964,2.0906656)(1.9880171,1.9147830790389142)(2.5080702,1.7389007)(2.6747394,1.9392601)(3.0908086,1.7385557)
\rput[bl](4.109475,1.844778){$M_0$}
\rput[bl](4.447253,2.3736668){$\F \subset \Lin(\H)$}
\psbezier[linecolor=black, linewidth=0.02](1.0108087,2.480778)(1.376924,2.5009513)(1.8323139,2.5047083)(2.1974754,2.440777911080255)(2.5626369,2.3768475)(2.9173744,2.2887836)(3.3352532,2.0674446)
\rput[bl](4.07392,1.4092224){$M_\tau$}
\psbezier[linecolor=black, linewidth=0.02, arrowsize=0.05291667cm 2.0,arrowlength=1.4,arrowinset=0.0]{<-}(3.384142,2.0452223)(3.6311882,1.9118891)(3.846798,2.0185556)(4.0285864,2.018555688858032)
\psbezier[linecolor=black, linewidth=0.02, arrowsize=0.05291667cm 2.0,arrowlength=1.4,arrowinset=0.0]{<-}(3.264142,1.871889)(3.4908087,1.8007779)(3.815253,1.5741112)(4.0019197,1.59633346663581)
\end{pspicture}
}
\caption{A quantum space-time}
\label{figqg}
\end{figure}
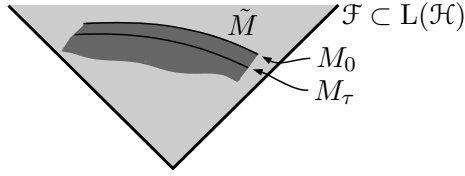

In other words, the space-time~$\tilde{M}$
is no longer described by a single metric, but by the whole family of metrics~$(g_\tau)$.
The situation becomes even more interesting if an interaction 
(as described by the causal action principle) is taken into account.
Then the different ``copies''~$M_\tau$ of space-time interact with each other
(as made precise by the constructions in~\cite{perturb}), giving rise to a complicated measure
which can no longer be described in terms of a family of classical metrics.
But, as shown in~\cite{perturb}, the resulting measure can
be described in terms of quantum fields.
This justifies the name ``quantum space time.''

\section{Connection to Other Approaches and Further Structures} \label{secoutlook}
\subsection{Connection to Lorentzian Length Spaces} \label{seclls}
A recent approach to non-smooth Lorentzian geometry is provided
by Lorentzian length spaces~\cite{samann}. We now explain how a causal fermion
system encodes the structure of a Lorentzian length space.
Our construction involves the choice of two length scales~$\ell_{\min}$ and~$\ell_{\max}$.
This is a subtle point which we explain in detail.
To this end, we return to the setting of a globally hyperbolic space-time~$(\scrM, g)$
and consider a  ``discrete timelike path'' \eqref{pndef} of space-time points~$p_0, \ldots, p_N \in \scrM$,
such that the Lorentzian distance of adjacent points is of the order~$\ell>0$.
In order to get a good approximation of a continuous path, the points should be sufficiently
close to each other. The relevant length scale~$\ell_{\max}$ must be chosen so small that
curvature effects are not relevant on this scale. Moreover, we choose~$\ell_{\max}$ so small
that the oscillations of the Dirac wave functions due to the rest mass do not come into play.
Thus we choose~$\ell_{\max}$ to be smaller than the Compton scale~$m^{-1}$.
The length scale~$\ell_{\min}$, on the other hand, must be chosen so large that
the effects of the regularization on the scale~$\varepsilon$ do not yet come into play
(where~$\varepsilon$ can again be thought of as the Planck length).
We thus obtain the admissible range
\beq \label{admissible}
\varepsilon \ll \ell_{\min} \;<\; \ell \;<\; \ll \ell_{\max} \ll m^{-1} \:,
\eeq
leaving us with a lot of freedom to choose the parameters~$\ell_{\min}$
and~$\ell_{\max}$ (note that for electrons, $m \varepsilon \lesssim 10^{-23}$).
If the Lorentzian geodesic distance~$\ell$ is within the range~\eqref{admissible},
then the kernel of the fermionic projector of the corresponding causal fermion system
is well-approximated by the unregularized massless kernel in Minkowski space
(as computed in detail in~\cite[Section~1.2]{cfs}).
In particular, the eigenvalues of~$xy$ scale like~$\ell^{-6}$.
Since these eigenvalues are defined in terms of objects of the corresponding
causal fermion system, we can generalize the notion the Lorentzian distance as follows.

Given a causal fermion system~$(\H, \F, \rho)$, we introduce the functional
\[ \ell \::\: M \times M \rightarrow \R^+_0 \:,\qquad \ell(x,y) = \left\{ \begin{array}{cc} |xy|^{-\frac{1}{6}}
& \text{if } \ell_{\min} < |xy|^{-\frac{1}{6}} < \ell_{\max} \\[0.3em]
0 & \text{otherwise} \:. \end{array} \right. \]
A finite sequence of points~$(x_0, \ldots, x_N) \in M$ is called a {\em{causal chain}} if for all~$n=1,\ldots, N$
the following conditions hold:
\begin{itemize}[leftmargin=3em]
\item[(i)] $x_n$ and~$x_{n-1}$ are timelike separated
\item[(ii)] $x_n$ lies in the future of~$x_{n-1}$
\item[(iii)] $\ell(x_n, x_{n+1}) >0$
\end{itemize}
The condition~(i) could be strengthened by demanding that~$x_n$ and~$x_{n+1}$
are spin-connectable (see~\cite[Definition~3.17]{lqg}).
The causal chain is said to {\em{connect}} the space-time point~$x$ with~$y$ if~$x_0=x$ and~$x_N=y$.
The {\em{length}} of the causal chain is defined by
\[ L[x_0, \ldots, x_N] = \sum_{n=1}^N \ell \big(x_{n-1}, x_n \big) \:. \]
The {\em{Lorentzian distance}} of two points is defined by
\beq \label{ddef}
\begin{split}
d(x,y) = \sup \big\{ &L[x_0, \ldots, x_N] \;\big|\; \\
&\quad \text{$(x_0, \ldots, x_N)$ is a causal chain connecting~$x$ with~$y$} \big\}\:.
\end{split}
\eeq
If the set on the right is empty, then the supremum is defined to be zero.
We also point out that the last supremum could be infinite, even if the points~$x$ and~$y$ are
close together. A typical example when this happens is when there
are closed causal chains (corresponding to closed timelike curves on a Lorentzian manifold).

With the above definitions,
a causal fermion system indeed gives rise to a Lorentzian length space
(our setting agrees precisely with~\cite{samann},
up to a few additional technical assumptions in~\cite{samann} which would have to be
verified in the applications).

\subsection{Connection to Causal Sets and Lattices}
Another approach to non-regular geometry and quantum geometry
are {\em{causal sets}} (see for example~\cite{sorkin}),
where the main structure is a transitive partial order relation~$\leq$
with the interpretation of ``lies in the future of.''
Given a causal fermion system, such a partial order relation can be
defined by
\[ x \leq y \qquad \text{if $x=y$ or~$d(x,y) > 0$} \]
(where~$d(x,y)$ is again the Lorentzian distance~\eqref{ddef}).
By construction, this relation is reflexive and transitive.
However, it is not necessarily anti-symmetric in the sense that~$x \leq y$ and~$y \leq x$
implies~$x=y$. Indeed, the above partial order might be trivial in the sense
that it holds for all pairs~$x,y \in M$. This is the case in particular if
the causal fermion system is constructed starting from a Lorentzian manifold
with closed timelike curves.

Having the partial order relation, one can follow the constructions in~\cite{cegla}
and introduce an {\em{orthogonality relation}} by
\[ x \perp y \qquad \text{if~$x \not \leq y$ and $y \not \leq x$} \]
and setting
\[ A^\perp = \{ x \in M \:|\: x \perp a \text{ for all~$a \in A$} \} \:. \]
Likewise, the sets of subsets of~$M$
\[ \L(M, \perp) := \{ A \subset M \:|\: A^{\perp \perp} = A\} \]
forms a {\em{lattice}}.

\subsection{A Riemannian Metric on~$\F$}
We finally introduce another inherent structure which has not been used so far
but which might be helpful or of interest for geometers.
To this end, let us again assume that our causal fermion system~$(\H, \F, \rho)$
is regular (see Definition~\ref{defregular}). Then we can also redefine~$\F$ as
the set of all self-adjoint operators on~$\H$ which have
exactly~$n$ positive and exactly~$n$ negative eigenvalues.
This has the advantage that~$\F$ has a smooth structure.
In order to keep the setting reasonably simple, we restrict attention to the
case that~$\H$ is finite-dimensional, in which case~$\F$ is a smooth manifold
(in the infinite-dimensional setting, $\F$ would be a Banach manifold).
Next, on~$\F$ the Hilbert-Schmidt norm gives a distance function
\[ d \::\: \F \times \F \rightarrow \R^+_0\:,\qquad
d(x,y) = \|x-y\|_{\text{\tiny{HS}}} := \sqrt{\tr \big((x-y)^2 \big)} \]
(note that the existence of the trace is not an issue even in the infinite-dimensional setting
because all operators in~$\F$ have finite rank).
The square of this distance function is smooth. Moreover, its first derivative vanishes
on the diagonal, i.e.~$D(d(x,.)^2)|_x=0$.
Therefore, taking its quadratic Taylor expansion
about a point~$x \in M$ gives a scalar product on~$T_x\F$, i.e.
\[ h_x \::\: T_x\F \times T_x\F \rightarrow \R \:,\qquad
h_x(u,v) = \tr(u v) \:. \]
Clearly, this mapping depends smoothly on~$x$ and thus defines a 
{\em{Riemannian metric}} on~$\F$.

Since every manifold can be endowed with a Riemannian metric,
the only point of this construction is the fact that the above Riemannian metric
is {\em{canonical}} in the sense that there is a distinguished Riemannian metric.
This Riemannian metric could be useful for different purposes.
Just to give an example, we here mention that it can be used for
a {\em{gauge-fixing procedure}}: In computations (like the perturbation expansion in~\cite{perturb}),
one often needs to
expand functions on~$\F$ in a Taylor expansion about a point~$p \in \F$.
Such an expansion is performed in a chart.
The expansion coefficients of second and higher order
depend in a complicated way on the choice of the chart.
The freedom in choosing the charts includes
the local gauge freedom in physics (for details see~\cite[Section~6.2]{perturb}).
Using the Riemannian metric~$h$, the Taylor expansion can be performed
in the distinguished chart given by the exponential map
\[ \exp_x : \U \subset T_x\F \rightarrow \F \:. \]
In these charts, the gauge freedom is fixed completely.

\Thanks {{\em{Acknowledgments:}} I would like to thank the organizers of the
workshop ``Non-regular space-time geometry'' held in Florence, June 2017,
for the kind invitation and financial support. I am grateful to Valter Moretti and Clemens S\"amann
for helpful discussions.

%\bibliographystyle{amsplain}
%\bibliography{../../aarbeit/felix}
\providecommand{\bysame}{\leavevmode\hbox to3em{\hrulefill}\thinspace}
\providecommand{\MR}{\relax\ifhmode\unskip\space\fi MR }
\providecommand{\MRhref}[2]{%
  \href{http://www.ams.org/mathscinet-getitem?mr=#1}{#2}
}
\providecommand{\href}[2]{#2}

\end{document}